# Axial magnetic field injection into thick, imploding liners

P.-A. Gourdain[1], M. Adams[1], J. Davies[2], C. E. Seyler[3]

[1] Extreme State Physics Laboratory, Physics and Astronomy Department, University of Rochester, NY 14627

[2] Laboratory for Laser Energetics, University of Rochester, NY 14627

[3] Laboratory for Plasma Studies, Cornell University, Ithaca, NY 14850

**Abstract**: MagLIF is a new fusion concept using a Z-pinch implosion to reach thermonuclear fusion. At present, he implosion is driven by the Z-machine using 27 MA of electrical current with a rise time of 100 ns. Since the implosion time is long compared to the heat diffusion time, MagLIF requires an axial magnetic of 30T to reduce heat losses to the liner wall. This field is generated well before the current ramp starts and needs to penetrate the transmission lines of the pulsed-power generator, as well as the liner itself. Consequently, the axial field rise time must exceed tens of microseconds. Any coil capable submitted to such a field for that long a time is inevitably bulky. The space required to fit the coil near the liner is large, increasing the inductance of the load. In turn, the total current delivered to the load decreases since the voltage is limited by driver design. Yet, the large amount of current provided by the Z-machine can be used to produce the required 30 T, simply by tilting the return current posts surrounding the liner, eliminating the need for a separate coil. However, the problem now is the field penetration time, which is limited by the skin effect inside the liner wall. This paper shows that large current densities also generate resistivity gradients which forces a redistribution of the current inside the liner wall much faster than the skin time. Numerical simulations show that fields larger than 30T can diffuse across the liner wall in less than 70 ns. external coil can be eliminated and replaced by return current posts with minimal helicity.

## Introduction

The MagLIF concept[1] is an experimental platform developed at Sandia National Laboratories that may place pulsed-power drivers at the forefront of fusion research, together with NIF and ITER. This concept relies on the implosion of a DT-filled metal liner using 40MA of current to match Lawson's criterion[2]. The original idea calls for three key elements. First, the liner must be thick enough to avoid magnetic Rayleigh-Taylor instabilities which would puncture the liner and cause mixing of high-Z material with the DT fuel. Second, an axial magnetic field is required to quench electron thermal conduction to the liner wall. This magnetic field must be compressed by the imploding liner to keep the electrons magnetized during the implosion[3]. Finally, the fuel must be heated just before the implosion starts. A cold fuel would let the magnetic field diffuse too quickly to the liner wall. Generating an initial axial magnetic field of 30T is one of MagLIF biggest technical challenges. The Helmholtz coil required to generate this magnetic field takes valuable space near the load, increasing the load inductance. At fix voltage, the total current cannot be delivered to the load. Yet, it is relatively easy to generate 30T when mega-amperes of current are flowing through the load[4]. However, when this current is pulsed it can only penetrate the wall by a distance δ, the skin depth, given by

$$\delta = \sqrt{\frac{\eta}{\pi f \mu_0}}. \qquad (1)$$

δ is on the order of 50 µm for an aluminum liner at a frequency f of 2.5 MHz (i.e. we suppose the current rise of 100 ns being ¼ of a sine wave). MagLIF calls for a wall thickness on the order of 300 µm-thick to mitigate magnetic Rayleigh-Taylor instabilities. Thus, a magnetic field rising in 100 ns does not have the time to diffuse through the liner on the implosion time scale.

It can be argued that resistivity depends on the temperature of the liner and increasing the liner temperature might accelerate electrical current diffusion. To first order, the resistivity of aluminum depends weakly on temperature (4x10$^{-3}$ Ohm.m/K), a liner at 100,000K would have a skin depth of 1 mm. When the liner is ohmically heated, the current must diffuse first before resistivity can rise. As a result, the actual skin depth is much smaller. Since magnetic pressure increases with the current, the outer layers of the liner rapidly fall into the warm dense matter (WDM) regime and the temperature dependence must be revised. Further, since the physical properties of WDM are not well established, the physical dependence of resistivity as a function of temperature is based on empirical and experimental resistivity models (e.g. Refs 5 and 6). Transitions to liquid and gas phases, then to a plasma state, are also a big factor in the overall impact of density and temperature on resistivity. Other effects, like Hall and electron inertia also play an important role at the plasma vacuum interface. In these regimes, resistivity is an important factor in any form of energy transport and reducing the complex physics happening when large current densities are involved cannot be reduced to a simple skin effect. There is another effect that displaces currents inside the metal rather than allowing them to diffuse and is caused by resistivity gradients. When current densities are large enough to produce such gradients, it is possible to inject axial magnetic fields through a thick liner wall faster than diffusion would normally allow. If an axial field can be generated by currents flowing inside slanted posts in the return current path of the pulsed-power generator and the resistivity gradients are large enough, then azimuthal eddy currents can penetrate the liner much faster than the diffusion time, allowing the axial field to also penetrate the liner wall. The angle of the posts controls the strength of the axial field generated at the liner surface. Posts are preferred to windings since they generate smaller voltage drops from one conductor to another[7].

After this introduction, a simplified one-dimensional model illustrates how magnetic field injection inside a conductor is accelerated by electrical resistivity gradients. Then, we uses three-dimensional numerical simulations to show how this effect happens in a more realistic model.

## Time evolution of electrical currents in the presence of resistivity gradients

We start by looking at a simplified model to show that current penetration does not follow skin effects when materials are heated unevenly. We start with Faraday's law

$$-\frac{\partial \vec{B}}{\partial t} = \vec{\nabla} \times \vec{E} \tag{2}$$

and Ampere's law in the quasi-static regime

$$\vec{\nabla} \times \vec{B} = \mu_0 \vec{J}. \tag{3}$$

In the resistive limit of Ohm's law, the electric field is directly proportional to the current via the resistivity $\eta$

$$\vec{E} = \eta \vec{J}. \tag{4}$$

We suppose in the rest of the paper that quasi neutrality holds so that

$$\vec{\nabla}.\vec{J} = 0. \tag{5}$$

As we know, the resistivity is a function of temperature and density, primarily. We use energy conversation to close our set of equations, using a simplified version of conservation of internal energy in one dimension,

$$n(Z+1)e\frac{\partial T}{\partial t} = \vec{E}.\vec{J}. \tag{6}$$

n is the number density of the material, Z its ionization and T is the temperature in eV. We have ignored phase changes and heat capacity which usually affect the initial temperature rise. Eq. (6) constraints $\eta$ to be an explicit function of space and time. Using Eq. (2) and (4) we find the time evolution of the magnetic field to be

$$-\frac{\partial \vec{B}}{\partial t} = \eta \vec{\nabla} \times \vec{J} + \vec{\nabla}\eta \times \vec{J}. \tag{7}$$

The first term in the right-hand side of Eq. (7) corresponds to eddy currents, responsible for skin effects. The second term corresponds to the generation of magnetic fields caused by the redistribution of electrical currents due resistivity gradients. For instance, a high intensity laser can prompt the sudden formation of resistivity gradients which redistribute currents[8,9]. In the region where resistivity gradients exist, the electric field is not curl free. This curl caused the emergence of a magnetic field. The JxB force pushes currents in regions with lower resistivity. In a purely diffusive case, the regions with largest electron scattering cross-sections (high resistivity) dispatch more electrons in regions with lower scattering cross-sections (low resistivity). Since regions with lower scattering cross-section send less electrons back into regions with larger scattering cross-sections, the current flows from high to low resistivity regions. However, when electrons are physically slowed down by an emerging resistivity gradient, the nascent magnetic field physically pushes these electrons outside of the high resistivity region. Unlike diffusion, which involves passive processes, this effect is actively driven.

Taking the curl of Eq. (7) and using Eq. (3) together with Eq.(5) we obtain the time evolution of the current density inside the materials

$$\frac{\partial \vec{J}}{\partial t} - \frac{\eta}{\mu_0}\vec{\nabla}^2\vec{J} + 2(\vec{v}_\eta.\vec{\nabla})\vec{J} - \vec{\nabla}(\vec{v}_\eta.\vec{J}) + (\vec{\nabla}.\vec{v}_\eta)\vec{J} = 0. \tag{8}$$

where we replaced resistivity gradients with the effective velocity

$$\vec{v}_\eta = -\frac{1}{\mu_0}\vec{\nabla}\eta. \qquad (9)$$

The second term in Eq. (8) corresponds to the usual current diffusion (i.e. skin effect). The third term corresponds to the advection of currents from a virtual "flow" with velocity $v_\eta$, given in Eq. (9) This velocity field is curl-free since it derives from a gradient. This term is responsible for the redistribution (i.e. advection) of electrical currents away from regions of high resistivity (as it would happen inside a circuit where resistors of decreasing resistance are successively connected in parallel). The fourth term forces current damping in the direction along resistivity gradients (as it would happen inside a circuit with resistors in series). The last term corresponds to current accumulation (rather than advection) in regions with lowest resistivity.

### Advection of currents caused by resistivity gradients

We further simplify our model to illustrate how current penetration evolves inside a motionless liner when resistivity gradients are present. We restrict our liner to a one-dimensional slab with infinite thickness beyond x=0 along the x-axis and infinite size in the y and z directions. We discuss here the time evolution of currents flowing perpendicular to the liner surface along the y-axis, so that $\vec{v}_\eta \cdot \vec{J} = 0$ in Eq. (8). If we take x as being the radial direction and y as being the axial direction, this model applies to the MagLIF geometry where curvature effects are neglected. Combing Eq. (2) to (4) we get the equation ruling the time evolution of the current in a slab model.

$$\frac{\partial J_y(x,t)}{\partial t} - \frac{1}{\mu_0}\frac{\partial^2[\eta(x,t)J_y(x,t)]}{\partial x^2} = 0 \qquad (10)$$

While Eq. (10) and the one-dimensional version of Eq. (8) are identical it is numerically advantageous to solve Eq. (10). We used the resistivity of aluminum in the WDM regime given by Faussurier et al.[10] (given in Figure 1-a). We also solved the usual diffusion equation:

$$\frac{\partial J_y(x,t)}{\partial t} - \frac{\eta(T,t)}{\mu_0}\frac{\partial^2 J_y(x,t)}{\partial x^2} = 0 \qquad (11)$$

This equation leads to formation of skin currents, giving rise to the formula given in Eq. (1). We are driving the edge current density at the left boundary in such a way that the total current inside the liner matches a typical Z current rise, shown in Figure 1-b. Eqs. (10) and (11) are solved using a spline-based finite difference scheme (e.g. Ref. 11). Both equations start with $J_y = 0$, the temperature and resistivity are set to room temperature values. We verified the correctness of the solution at t=100ns by comparing the magnetic field time-integrated throughout the whole numerical simulation from Eq. (2), against the magnetic field space-integrated at the end of the computation, as given by Eq. (3). The difference between the two is less than a percent and for all cases discussed in this section. It is important to note that the simple model presented here does not claim to capture all physical aspects an imploding liner which current densities are large enough to caused sizable temperature gradients inside the material. Realistically, several other effects, such as melt, vaporization or ablation, can enhance or inhibit the evolution of the current.

Figure 2 shows the time evolution of the current density (Figure 2-a), resistivity (Figure 2-b) and temperature (Figure 2-c) versus liner depth for the resistivity model plotted in Figure 1-a. The liner thickness is large compared to the penetration of the current, avoiding to deal with boundary conditions at the right boundary. The color scale for each profile corresponds to a time spanning 100ns, in 10ns increments. Initially (t<30ns), resistivity gradients are negligible and the current penetration is diffusive, following the conduction Eq. (11). As soon as a resistivity gradient starts to form, even in its mildest form (t=20ns), the current shape radially changes. This gradient is created by the temperature increase due to Ohmic heating, visible in Figure 2-c. At this point in time, the current penetration trades its diffusive features for advection-like characteristics, following Eq. (8). Since the resistivity gradient points outwards, the advection velocity $\vec{v}_\eta$ points inwards. We are now in the advection phase where the leading edge of the current is start to move (rather than diffuse) inwards (t=40ns).

Besides uncertainties inherent to the warm dense matter resistivity model used here, two important effects have been ignored in this simulation. The most important effect is material ablation caused by Ohmic heating, localized to the outermost layers of the liner. The innermost layers of the liner are surrounded by dense matter and transition to a warm dense matter state instead. In this case, we do not expect drastic material expansion and the proposed model should be relevant there. On the outer edge, plasma will surely form and we can expect a drop in resistivity. If the resistivity is low enough, resistivity gradients can point radially inward on the outer surface and current advection will reverse. At this point the current already injected inside the liner will be cut-off from the current source located at the outer edge. Yet, any current already in the material will behave similarly to the current of in Figure 2. The starved currents will certainly not produce such large resistivity gradients and we can expect a reduced advection speed. Physically, the current will reach the inner edge of the liner and ablation will also occur there. The ablated plasma will generate resistivity gradient that will enhance current penetration into the cold fuel. At the interface, fuel and liner material will surely mix, making the fuel harder to heat due to the presence of high Z impurities. However, this discussion is based on Spitzer resistivity and does not account for electron inertia. Electron inertia becomes important at the liner-vacuum interface and the effective resistivity is much larger at the interface than inside the material. Hence, the resistivity gradients on the outer and inner liner surfaces points away from the liner. So the advection forces current into the liner on the outer surface and also keeps the currents into the liner on the inner surface. This effect is discussed further in the next section. A secondary effect is capable of modifying the conclusion presented in this section. The model does not account for the actual motion of the liner as it implodes. Liner motion is important because it generate compression inside the material. As a result, both the temperature and the density rise inside the liner. Our model supposed the density constant throughout the current ramp. The next section also look at this effect using 3D simulations.

It is interesting to compare the purely diffusive case with the advection case. Figure 2 shows the current distribution at t=100ns, assuming a purely diffusive behavior, as given by Eq. (11). The difference between the diffusive current profile and the more realistic case, where the impact of resistive gradients is accounted for, is striking. The temperatures obtained in the purely diffusive case are an order of magnitude larger than in the advective case. Since the resistivity model used here has not been validated beyond 50 eV, it is preferable not to speculate further on the veracity of the numerical solution. Using the 1/e folding in current, we computed the advective skin depth at 400µm, when solving Eq. (10) numerically. The diffusive skin depth is 80µm, computed from Eq. (11), five times smaller than previously computed. This skin depth is only 60 % larger than the skin depth computed by using Eq. (1). For an aluminum liner at room temperature, the model given by Eq. (10) shows that the current has fully penetrated a 300µm-thick MagLIF liner after 80ns.

## Axial field injection into thick liners

The previous model shows that currents (hence magnetic fields) can penetrate an aluminum liner much faster than the skin time when the axial current density is large enough to generate strong resistivity gradients. Any means to generate an external magnetic field using the current available from the Z machine would simplify platform design and decrease load inductance since an axial field coil would become superfluous. For instance, Automag[12] uses a conductive helical path allowing the formation of axial magnetic fields directly inside the liner, eliminating the Helmholtz coil altogether. It was shown that the helical structure inside the liner does not impact noticeably magnetic Rayleigh-Taylor instabilities[12].

Another possibility is presented herein. We can use the return-current posts to force the return current to partially run in the azimuthal direction. Rather than creating a helical path with become rapidly too inductive, we propose to tilt the posts. The azimuthal current will generate a magnetic field that will penetrate the liner in concert with the axial current. Physically, eddy currents generated inside the liner by the time varying axial field will penetrate the liner. Once they cross the liner wall, the axial field will grow inside the liner cavity. While the previous model was simplified to highlight the essence of current advection, it has limited applicability to assess how much axial field can be injected through the liner. We used the code PERSEUS[13] to evaluate more precisely this effect and how this mechanism takes place when multiple resistive mechanisms are taken into consideration. The three-dimensional numerical simulation used here takes into account the phase transitions of aluminum as well as warm dense matter mixed with plasma and gas states[5]. Figure 3 shows the geometry used in the simulations. The domain footprint is 18mm long by 18mm wide. Its height is 11mm with a total number of 88 million grid cells (512x512x336). The geometrical resolution is 40 µm, which is ten times larger than the resolution of the one-dimensional model. The numerical simulation used 4096 processors. The liner is made of aluminum, with a wall thickness of 300 µm and a radius of 3mm. The wall is 8 grid cell across so the current advection is under-resolved. This lack of resolution is mitigated by the conservative algorithms built into PERSEUS, which are capable of tracking momentum and energy transport accurately even across few cells. For instance, PERSEUS can reproduce accurately the physics of under-resolved geometries[14]. All boundary conditions are open except at the anode-cathode gap (indicated by "A-K gap" in Figure 3) where we imposed a 1/r magnetic field decay, generated by a current rise from 0 to 27 MA. The current rise used is given by Figure 1-b. In this magnetic configuration, the current flows down the liner, from the anode to the cathode. The outer posts are slanted by 22.5° with the vertical. We stopped the simulation 70 ns into the current rise, when the outer surface of the liner starts to move. Our main interest here was not the full implosion but how much field can be injected inside the liner before laser pre-heating takes place. Once the plasma is heated by the laser, the magnetic field is frozen into the hot fuel and very little supplemental field can penetrate. We chose to simulate an empty liner to help focus the discussion on the interaction of the field with the liner wall rather than with the fuel. However, simulations not presented here showed that the presence of the fuel did not impact significantly the axial magnetic field penetration if the gas temperature is below one eV.

The major difference between the previous model and PERSEUS comes from the electric field model. Since PERSEUS is a two-fluid code, it uses the generalized Ohm's

$$\vec{E} = -\vec{u}\times\vec{B} + \eta\vec{J} + \frac{1}{en_e}\left[\vec{J}\times\vec{B} - \vec{\nabla}p_e\right] + \frac{m_e}{n_e e^2}\left[\frac{\partial \vec{J}}{\partial t} + \vec{\nabla}\cdot\left(\vec{u}\vec{J} + \vec{J}\vec{u} - \frac{1}{en_e}\vec{J}\vec{J}\right)\right] \quad (12)$$

instead of Eq. (4). We ignore electron pressure in the simulation presented here. Eq. (12) is used to advance the current density rather than the electric field. The time advance of the field is given by the generalized Ampere's law

$$\varepsilon_0\mu_0\frac{\partial \vec{E}}{\partial t} = \vec{\nabla}\times\vec{B} - \mu_0\vec{J}. \quad (13)$$

Eq. (12) affects the time evolution of the current through other dissipative effects which are not accounted for in the standard resistivity definition given by Eq. (4). To account for all terms in the generalized Ohm's law, we grouped all dissipative resistive terms from the code into an effective resistivity, given by

$$\eta_{\text{eff}} = \frac{\vec{E}\cdot\vec{J}}{J^2}. \quad (14)$$

While not dissipative, the Hall effect

$$\vec{E}_{\text{Hall}} = \frac{1}{en_e}\vec{J}\times\vec{B} \quad (15)$$

can enhance magnetic field penetration[15] when the characteristic scale length of the plasma is on the order of the inertial scale. Finally, the electron inertia, with electric field

$$\vec{E}_{\text{inertia}} = \frac{m_e}{n_e e^2} \left[ \frac{\partial \vec{J}}{\partial t} + \nabla \cdot \left( \vec{u}\vec{J} + \vec{J}\vec{u} - \frac{1}{en_e} \vec{J}\vec{J} \right) \right], \tag{16}$$

does limit electron flows at the plasma vacuum interface[16], and can be seen as a resistive effect. The sole purpose of Eq. (14) is to encapsulate all possible effects causing $\nabla \times \vec{E} \neq 0$ and causing the rise of induced magnetic fields, allowing to show in one plot where gradients are located. This choice does not play any role in numerical simulations.

Figure 4 shows a series of snapshots of the current density (plotted on the left) and the generalized resistivity (plotted on the right). Each frame is taken 10ns after the previous one, until the outer surface of the liner starts to move. While we expect a different trend between the full MHD simulations and the one-dimensional model discussed previously, both cases share similar features, namely resistivity gradients and a current peaking inside the liner wall rather than at the edge, a sign of diffusion. On the left side of Figure 4-a, we clearly see that the current is already flowing inside the liner wall, 20 ns into the current ramp. This rapid penetration already shows that the current distribution is not controlled by resistive diffusion in two-fluid MHD. This effect is caused by resistivity gradients coming from electron inertia at the edge of the liner. These effects, often found at the material-vacuum interface, effectively generates resistivity gradients which point away from the liner wall, both on the inner and the outer surfaces of the liner, thereby keeping the current into the wall. This increase in edge resistivity is visible on all right panels of Figure 4-a, where the resistivity is systematically higher on the inner and outer surfaces of the liner. This current profile does not fully agree with the current distribution presented in Figure 2, which cannot capture this effect. Figure 4-b shows the current density and the generalized resistivity 30ns into the current ramp. The current distribution clearly shows a peak at this time. The current peak is located 100µm from the outer surface of the liner. Resistive gradients are also visible. Figure 4-c shows that the current peak has moved radially inward after 10ns. The resistive gradients have steepened. The advection of the current continues as the peak meets the inner surface of the liner in Figure 4-d.

There is one notable difference between the 1D and 3D simulations. In the 3D simulation, the current peak reaches the edge of the liner early, compared to the simple model describe in the previous section. This difference is attributed to the existence of strong initial resistivity gradients at the material-vacuum interface present in the 3D simulation. As the result, the advection of the current in the 3D simulations gets an early start. However, they share two common features. First, both simulations clearly show a non-diffusive evolution of the current distribution. Second, the speed of the current peak in both simulation is the same. It takes 30 ns to cross 150µm in the 1D and 3D simulations. This is not surprising since the advection speed given by Eq. (9) is strictly connected to the resistivity gradients, which are similar in both simulations (on the order of 10mΩ). Figure 4 does not go beyond 50 ns since the outer surface of the liner start to move inward. At this point, motion generates a complex interaction between the current and the liner, which is far beyond the reach of the 1D model. Figure 4 might imply that the skin depth in the 3D simulation is even larger than in the 1D case. However, the resistivity gradients at the inner surface of the liner points radially outward due to the high resistivity at the plasma vacuum interface. These gradients keep the current inside the liner until ablation allow the plasma to expand inside the liner cavity. At this point, the current starts to flow in the liner cavity.

Eddy currents generated by the time evolution of the axial field have also reached the inner surface of the liner, allowing some portion of the axial field to penetrate that far. The plasma that expands in the liner cavity, carries with it the axial (and azimuthal) magnetic field. The axial magnetic field reaches the axis 60 ns into the current ramp. Figure 5 now shows the overall simulation domain, 70 ns into the current ramp, after the inner liner wall has moved inward by 300 µm. All plots on Figure 5 are on the logarithmic scale. According to Figure 5-a, ablated plasma from the inner surface of the wall has reached the axis. This plasma has a low density ($<5 \times 10^{26} m^{-3}$) but a relatively high temperature (T>500 eV). This pressure is relatively large compared to atmospheric pressure, which corresponds to the initial fuel pressure in MagLIF. While most of the current is still locked inside the liner wall, Figure 5-b shows that a good fraction of this current has reached the axis. A portion of that current is the induced (eddy) azimuthal current. As the azimuthal current disappear on axis, it allows the axial magnetic field to grow inside the liner well above 30T and without compression from the imploding liner. Figure 5-b shows that the axial field can reach up to 200 T from compression by the ablated plasma blown off the inner surface of the liner wall.

## Conclusion

We have shown using a very simple model that current advection due to resistivity gradients is an important mechanism of current penetration inside materials that are not heated homogeneously. This is particularly true when materials are pushed into the warm dense matter regime. In this context, the main difference between advection and diffusion is the unidirectionality of advection, which enhances current penetration along the direction antiparallel to resistivity gradients. When mild gradients are present (~0.5Ω), current penetration is visibly altered and the time evolution of the current changes dramatically. However, the actual skin depth as given by Eq. (1) approximates well the depth at which the current flows inside the material. When gradients become large (>5Ω), Eq. (1) is not even a good approximation of the current penetration depth, which can be an order of magnitude larger. Since both effects are strongly coupled to material response to heat, the effects discussed here should be quantified rather than invoking skin depth arguments.

This is particularly true for an imploding MagLIF liner on the Z-machine, where current densities are so large that they can trigger this type of current penetration. This can be an issue if the current reaches the inner surface of the liner, where it will cause plasma ablation, mixing the high-Z material from the wall with the low-Z fuel. But we can turn this mechanism to our advantage as we can now inject an eternal magnetic field inside a liner on time time-scale much smaller than the skin effect time scale. Instead of using a Hemlholtz coil pair to pre-magnetize the fuel, one can use slanted return posts to produce the necessary axial field to magnetize the fuel during the initial current ramp. Since the current flows across the liner much faster than diffusion would normally allow, the axial field can penetrate across the liner wall before the inner surface of the liner has started to move. By the time laser heating takes place, the required pre-magnetization field would have reached the axis. Under such conditions, the magnetic topology would be similar to the standard MagLIF scheme. The beneficial effects discussed herein would add to a substantial reduction of magnetic Rayleigh-Taylor instabilities caused by twisted return current paths[17].

**Acknowledgements**: This research was partially supported by the Department of Energy National Nuclear Security Administration under Award Numbers DE-SC0016252 and DE-NA0001944.

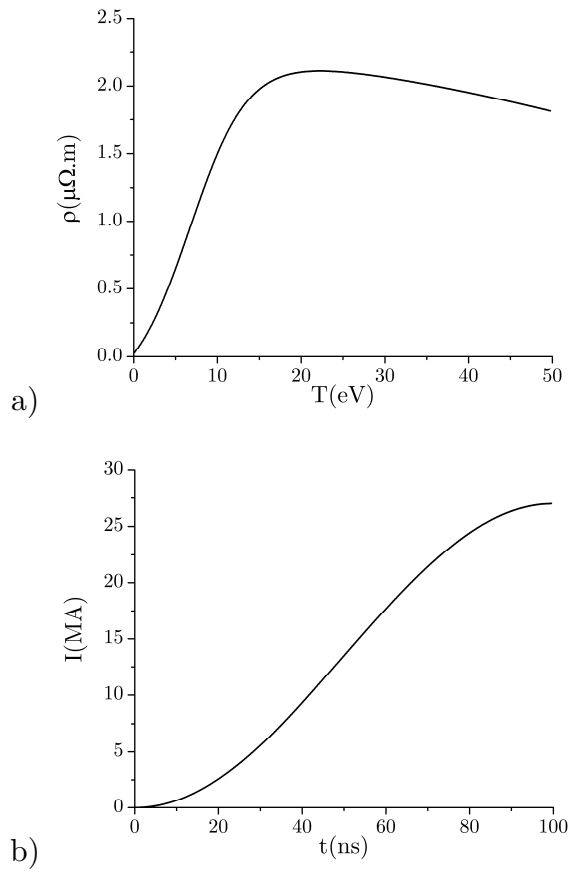

**Figure 1. a) Resistivity model as a function of temperature. b) The current ramp imposed as boundary condition for Eqs. (10) and (11)**

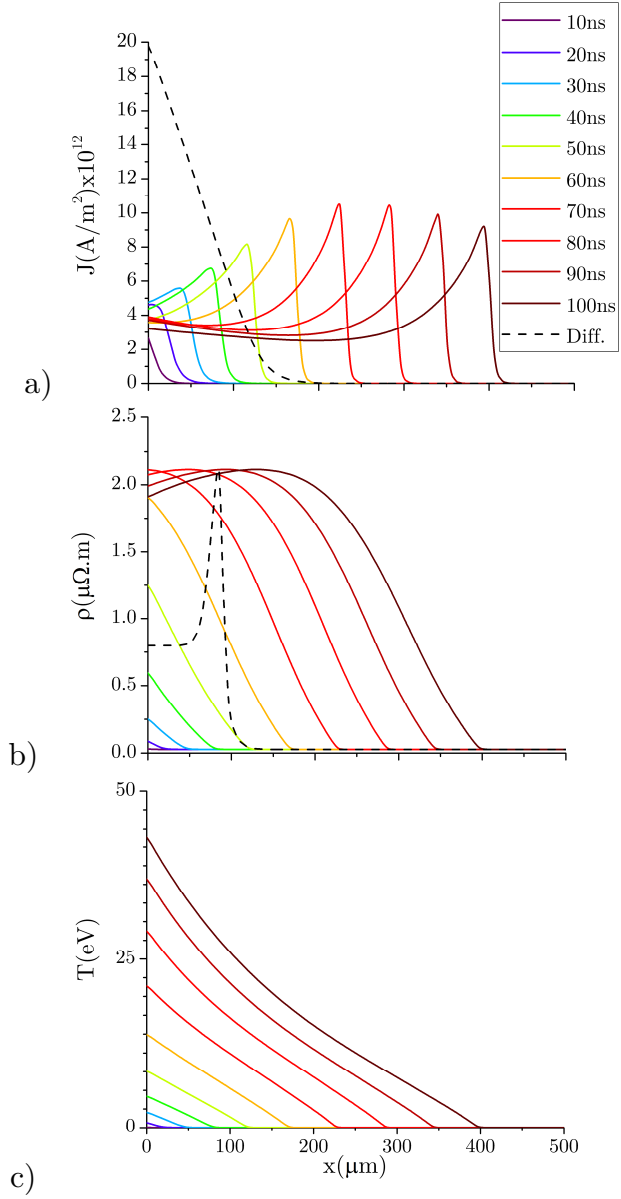

**Figure 2.** Time evolution of the a) current density, b) resistivity and c) temperature from 0 to 100 ns in 10 ns increments, solving Eq. (10) using the resistivity shown in Figure 1. The black dashed line corresponds to the current density and resistivity for the diffusive model given by Eq. (11) at t=100ns for current evolution given in Figure 1-b.

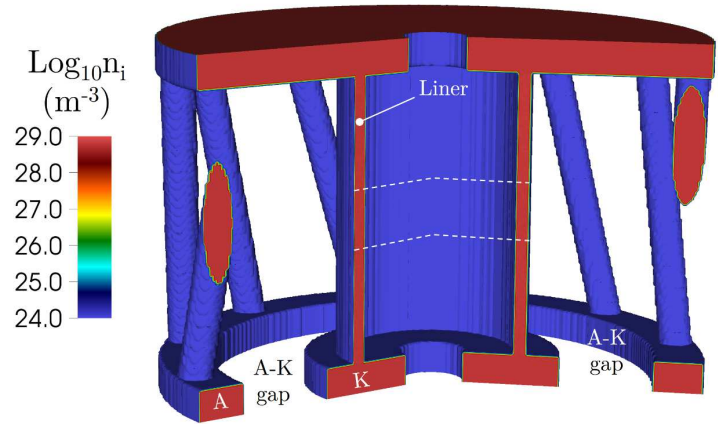

**Figure 3.** Overall geometry used in the simulation showing the liner, the anode (A) the cathode (K) and the A-K gap. The dashed lines indicate the section plotted in Figure 4.

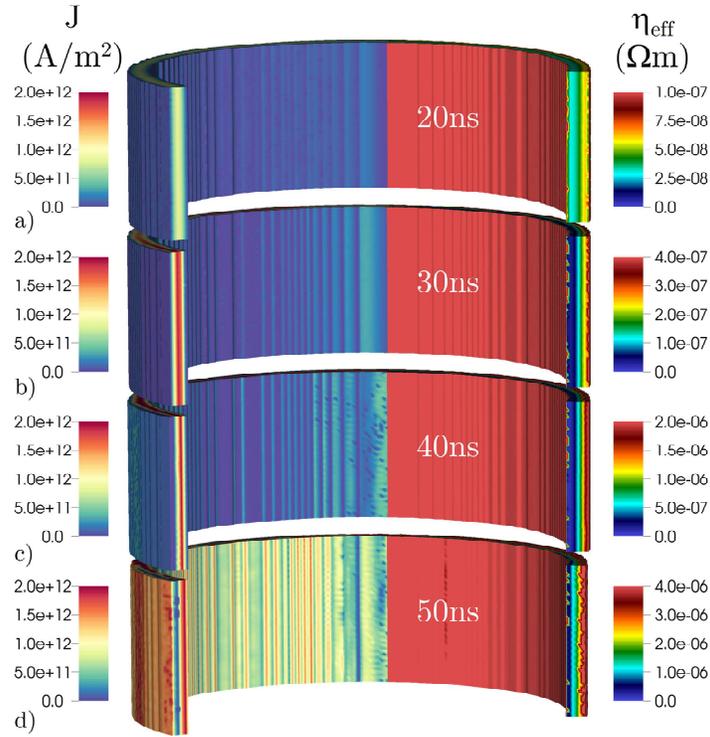

**Figure 4.** Time evolution of the liner for a) 20ns, b) 30ns, b) 40ns, and d) 50ns. Current density (left) and effective resistivity (right) are plotted on the linear scale. The resistivity color scale is adjusted for each panel. The current density color scale stays the same. Regions where the ion density is below $6 \times 10^{28} m^{-3}$ are not displayed.

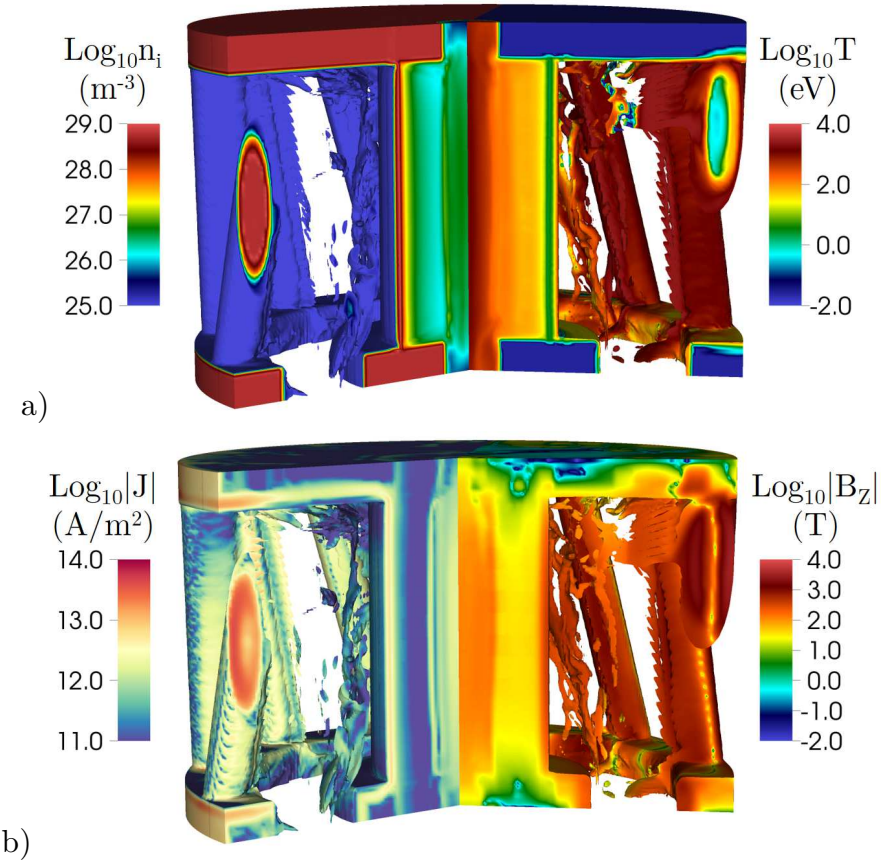

Figure 5. Liner implosion shown when the liner starts to move, 30 ns before current peak. The simulation results show a) the ion density (left) and the temperature (right) on the $\log_{10}$ scale. b) The current density and the magnetic field, also on the $\log_{10}$ scale are shown. Regions where the ion density is below $10^{25}$m$^{-3}$ are not displayed.